\documentstyle[eqsecnum, epsf, aps]{revtex}

\newcommand{\f}{\varphi}
\newcommand{\C}{V(\f)}

\begin{document}

\draft

\title{ Global Dynamics of Cosmological Expansion with a Minimally Coupled 
Scalar Field }

\author{ David I. Santiago\thanks{david@spacetime.stanford.edu}}
\address{ Department of Physics, Stanford University }
\author{ Alexander S. Silbergleit\thanks{gleit@relgyro.stanford.edu}} 
\address{ Gravity Probe-B, W.W.Hansen Experimental Physics Laboratory, 
Stanford University}

\date{\today}

\maketitle

\begin{abstract}
We give a complete description of the asymptotic behavior of a 
Friedmann-Robertson-Walker Universe with ``normal'' matter and a minimally 
coupled scalar field. We classify the conditions under which the Universe is or
is not accelerating. In particular, we show that only two types of large time 
behavior exist: an exponential regime, and a subexponential expansion with the 
logarithmic derivative of the scale factor tending to zero. In the case of the 
subexponetial expansion the Universe accelerates when the scalar field energy 
density is dominant and the potential behaves in a specified manner, or if 
matter violates the strong energy condition $\rho + 3p >0$. When the expansion
is exponential the Universe accelerates, and the scalar field energy density is
dominant. We  find that for the Big Bang to occur at zero scale factor, the 
equation of state of matter needs to satisfy certain restrictions at large 
densities. Similarly, a never ending expansion of the Universe constrains
the equation of state at small matter densities.
\end{abstract}


In this letter we give a complete generic description of the late time 
asymptotic behavior of the cosmological expansion under quite general 
conditions. While there is literature devoted to specific potentials and 
interesting attractor solutions within those models \cite{q1,peb2,q2}, 
in the present work we study the \emph{generic} features of the late time 
asymptotics for \emph{any} potential. We choose units in which $G=c=1$.

Recent results from supernovae type Ia \cite{sn1a1} indicate that the
Universe may be accelerating. If this is the case, we are currently entering a 
period of cosmological inflation. An accelerating Universe is sometimes 
interpreted as evidence for a cosmological constant, which is not necessarily 
so. Rather, this testifies only that the dominant material in the Universe is
characterized by an equation of state that satisfies $\rho + 3p < 0$
~\cite{peb}. ``Exotic'' matter described by this type of equation of state has 
been called quintessence \cite{q1}. A self-interacting scalar field is one way 
to provide acceleration without assuming this peculiar property of "normal" 
matter \cite{q1,peb2,q2}. As far back as ten years ago, Ratra and Peebles
considered a cosmology where the scalar field energy density becomes dominant 
at the present cosmological epoch \cite{peb2}. They found asymptotically stable
equilibrium solutions in which the scalar field energy density dominates the 
dynamics. Their work has recently been resurrected by Liddle and Scherrer 
\cite{q2}. These results correspond to certain special forms of the scalar 
field potential and also rest upon other assumptions. 

Since the Universe is very homogeneous and isotropic on large scales, it
can be
approximated by a Friedmann-Robertson-Walker metric:
\begin{equation}
ds^2= - dt^2 + R^2(t)\left( \frac{dr^2}{1-kr^2} + r^2d\Omega^2\right) \ ,
\end{equation}
where $d\Omega$ is the element of solid angle and $k=$ -1, 0, or 1
according to
whether the Universe is open, flat, or closed. In this work we consider mainly
open and flat cosmologies. The only energy-momentum tensor consistent with
homogeneity and isotropy is the one corresponding to a perfect fluid. Therefore 
we consider a Universe filled with material of energy density $\rho$ and
pressure 
$p$, and with a self-interacting scalar field $\f$ that has potential $\C$.
We 
assume that
\begin{eqnarray}
\rho > 0 \label{eq:r} \\
\rho + p > 0 \label{eq:rp} \\
\C \geq 0 \label{eq:lamb}
\end{eqnarray}
Those are fairly reasonable restrictions. Condition (\ref{eq:r}) states that 
the energy density is positive, while (\ref{eq:lamb}) ensures that the scalar
field energy density never becomes negative. We believe it is 
unreasonable to consider material with pressure more negative than the vacuum 
($p= -\rho $) and therefore we impose (\ref{eq:rp}). We also exclude the 
vacuum case itself because the cosmological constant can be treated as part of
the scalar field potential. The matter $\rho$ in the Universe is made up of 
baryons, photons, neutrinos, dark matter, and whatever else it might be. We 
suppose the pressure to be some function of the density, $p=p(\rho)$. The 
equation of state in the present epoch is very close to that of a pressureless
fluid (``dust''), but we consider an entirely general equation of state, 
assuming only (\ref{eq:rp}) and 
\begin{equation}
p(0)=0 \label{eq:p0}
\end{equation}
(in the absence of matter there should be no pressure at all).

The field equations are
\begin{eqnarray}
3 \frac{\dot{R}^2}{R^2}= 8 \pi  \rho + \dot{\f}^2 + \C - 3 \frac{k}{R^2}
\equiv
\rho_T- 3 \frac{k}{R^2}  
\label{eq:h} \\
\ddot{\f} + 3 \frac{\dot{R}}{R} \dot{\f} + \frac{1}{2} \frac{d\C}{d\f}=0 
\label{eq:f} \\
\dot{\rho} = -\frac{3 \dot{R}}{R} (\rho + p) \label{eq:e}
\end{eqnarray}
where the total energy density $\rho_T = 8 \pi  \rho +\rho_\f$, and $\rho_\f =
\dot{\f}^2 + \C$ is the energy density of the scalar field. The equations also
combine to give
\begin{eqnarray}
3\frac{\ddot{R}}{R}=-4\pi\left(\rho+3p\right)+\C-2\dot{\f}^2\equiv \nonumber \\
-4\pi\left(\rho+3p\right)-\frac{1}{2}\left(\rho_\f+3p_\f\right), \label{eq:ac}
\end{eqnarray}
where $p_\f = \dot{\f}^2-\C$ is the scalar field pressure. Thus the expansion 
is accelerating when $\left(\rho_\f+3p_\f\right)<0$ and larger in magnitude 
than $8\pi\left(\rho+3p\right)$.

We consider an expanding (as opposed to contracting) Universe. Hence we choose
the positive square root in equation (\ref{eq:h}), which, in view of equations
(\ref{eq:r}), (\ref{eq:lamb}) and $k=0,-1$, implies $\dot{R}/R > 0$; hence 
$\dot{R} > 0$, so $R \rightarrow \infty$ as  $t \rightarrow\infty$. We thus 
have a 4D dynamical system  whose trajectories $\{R(t),\,\rho(t),\,\f(t),\,
\dot{\f}(t)\}$ are specified by initial values $R_0>0,\,\rho_0>0,\,\f_0,\,
\dot{\f}_0$ at some $t=t_0$.

The energy conservation equation (\ref{eq:e}) can be integrated to
\begin{equation}
\int_{\rho}^{\rho_0} \frac{d\xi}{\xi + p(\xi)} = 3 \ln \left ( \frac{R}{R_0}
\right ) \label{eq:int}
\end{equation}
By equations (\ref{eq:e}) and (\ref{eq:rp}), $\rho(t)$ is a monotonically 
decreasing (positive) function, therefore it has a nonegative limit when $t
\rightarrow \infty$. However, for large times $R \rightarrow \infty$, so the 
integral in the left-hand side of (\ref{eq:int}) must diverge. Because of our 
conditions (\ref{eq:rp}) and (\ref{eq:p0}), this can only happen if $\lim_{t 
\rightarrow \infty}\rho(t)=0$; the matter density in the Universe goes 
monotonically to zero with its expansion. Note that the divergence of the 
integral (0.10) at $\rho=0$ requires the pressure go to zero  fast enough for 
small densities (c. f. the integral converges if $p \propto \rho^\gamma$, $
\gamma<1$ as $\rho \rightarrow+0$). It is sufficient for the divergence that 
the derivative $p'(0)$ exists, which we assume in the sequel.

We now briefly turn to the Big Bang. We require that $R_0\rightarrow 0$ at 
the time (either finite or $-\infty$) of the BB singularity in the past. As 
$R_0 \rightarrow 0$,  $\ln \ (R/ R_0) \rightarrow \infty$, and again the 
integral in (\ref{eq:int}) must diverge, which is impossible unless $\rho_o  
\rightarrow \infty$ (a definition of the BB singularity). However, this 
integral does not diverge at $\rho_0 = \infty$ for any thinkable dependence 
$p(\rho)$: for the divergence, the pressure must not grow too fast at large 
densities. Thus a surprising fact is that the very existence of the BB imposes 
a limitation on the equation of state in the early Universe. In terms of a 
power scale, if $p \propto \rho^\gamma$ for $\rho\gg1$, then $\gamma\leq1$. It 
is suggestive that if $\gamma>1$ not only does the integral converge, but 
causality is violated: the speed of sound is 
$$
c_s^2=\left( \frac{\partial p}{\partial \rho} \right)_S \propto 
\gamma \rho^{\gamma-1} \, ,
$$
so it becomes greater than one for sufficiently large $\rho$. 

Once again, for the BB to occur at $R_0=0$ and/or the infinite expansion to 
exist,the equation of state must ensure that the integral (\ref{eq:int}) 
diverges at $\rho_0 = \infty$ and/or $\rho=0$. This result does not depend on 
the scalar field system, as it is minimally coupled; in particular, it is valid
for $V\equiv0$ as well.

While certainly thought provoking, these considerations about the BB should be 
taken with caution as the situation in the very early universe is probably 
considerably more complicated. In fact, at very early times the universe 
probably underwent a period of inflation \cite{peb,kt,lind}. In such a period 
the evolution of the universe is not dominated by the energy density of regular 
matter, but by the energy density of the inflation field(s). In fact 
inflationary scenarios in which the expansion is exponential violate the weak 
energy condition (\ref{eq:rp}). Since inflation is the only known solution to 
the horizon problem \cite{peb,kt,lind}, our assumptions are probably too 
simplistic when applied to an early universe which underwent an exponential 
inflationary phase. On the other hand, accelerated sub-exponential expansion 
that solves the horizon problem can occur in the early universe and condition 
(\ref{eq:rp}) can be satisfied. Global constraints regarding the weak energy 
condition and generic inflationary scenarios have recently been studied by 
Vachaspati and Trodden \cite{inf}.Whether the existence of the BB at $R_0 = 0$ 
implies an equation of state which is causal is a matter of a separate 
investigation. 

In order to study the limit of large times for the cosmological dynamical 
system (\ref{eq:h})---(\ref{eq:e}), we first note that equations (\ref{eq:e}) 
and (\ref{eq:f}) combine into
\begin{equation}
\dot{\rho_T} = -\frac{6 \dot{R}}{R}\left[4 \pi \left(p+\rho\right)+\dot{\f}^2
\right]<0,
\end{equation}
with the inequality implied by (\ref{eq:rp}) (it is straightforward to see 
that the density does not turn to zero at any finite time). This shows that 
the initial store of energy is dissipated into the expansion, and that, just  
as $\rho$, the total energy density $\rho_T$ is a (positive) decreasing 
function of time; it is a Lyapunov function of our 
dynamical system (e. g. ref~\cite{rei}, 2.3). Thus $\rho_T$ has a nonegative 
limit $\lim_{t \rightarrow \infty} \rho_T=\rho_T^{\infty}\equiv\rho_T^{\infty}
(R_0,
\rho_0,\f_0,\dot{\f}_0) \geq0$; since $\rho\rightarrow0$, we have also $
\lim_{t \rightarrow \infty}\rho_\f=\rho_T^{\infty}$. Therefore the large time 
behavior of the cosmological expansion is described by
\begin{eqnarray}
3 \frac{\dot{R}^2}{R^2}= \rho_T^{\infty}+\dots \label{eq:ha} \\
\rho\rightarrow0 \label{eq:ra} \\
\rho_{\f}= \dot{\f}^2 + \C= \rho_T^{\infty}+\dots \label{eq:rfa} \, ,
\end{eqnarray}
where dots stand for the terms tending to zero; in addition, equations 
(\ref{eq:f}) and (\ref{eq:ac}) become
\begin{equation}
\ddot{\f} + \beta \dot{\f} + \frac{1}{2} \frac{d\C}{d\f}+\dots=0,\qquad \beta
\equiv\sqrt{3\rho_T^{\infty}}\geq0 
\label{eq:fa}
\end{equation}
\begin{equation}
3\frac{\ddot{R}}{R}=\rho_T^{\infty}-3\dot{\f}^2+\dots=3\C-2\rho_T^{\infty}+
\dots \label{eq:aca}
\end{equation} 
Depending on whether $\rho_T^{\infty}$ is positive, there are clearly two 
different large time regimes of the expansion. 

If $\rho_T^{\infty} > 0$, the Universe expands (and of course accelerates)
exponentially; this is the only possible regime when the potential is bounded 
away from zero, $\C\geq V_{min}>0$, for instance, if a cosmological 
constant exists.  

The expansion is subexponential with $\dot{R}/R\rightarrow0$ when $
\rho_T^{\infty} = 0$, which requires the zero value to belong to the closure 
of the $\C$ image. In this case, the Universe could accelerate or decelerate 
depending on further properties of the scalar field potential.

We first consider the subexponential case, i. e., any solution with $
\rho_T^{\infty} = 0$. 

Obviously, here $\lim_{t \rightarrow \infty} \dot{\f}(t)=0$, $\lim_{t
\rightarrow \infty} V(\f(t))=0$, so that the scalar field also goes to 
some limit $\f_{\infty}$, either finite or infinite, and $V(\f_\infty)=0
$. If the limit is finite, $|\f_{\infty}|<+\infty$, then, just by inequality
(\ref{eq:lamb}), $d \C/ d \f |_{\f=\f_{\infty}}=0$, $d^2 \C / d \f^2 |_{\f=
\f_{\infty}} >0$; the system evolves towards one of its finite stable critical
(fixed) points with the zero minimum of the potential. It is not 
difficult to show that the expansion in this case is not accelerating unless 
$p'(0)<-1/3$, which means $\rho + 3p < 0$ at least for small enough 
densities; the scalar field thus cannot "outweigh" the "normal" matter under 
such conditions.

However, if $\f_{\infty}=\pm\infty$, which requires $\lim_{\f\rightarrow\pm 
\infty}\C=0$, the situation with the acceleration  can be different depending 
on the details of behavior of the potential at infinity (the system in any 
case goes again to a fixed point, only corresponding to an infinite value of 
the field). We describe here two classes of potentials allowing the 
acceleration.

a) {\it Potentials vanishing exponentially at infinity}. For some $a> 1$, let
$\C = a(3a - 1) \exp{(-2\f /\sqrt{a})}\left[1+{\rm o}(1)\right]$
 as $\f\rightarrow +\infty$, and this asymptotic formula may be differentiated.
If at large times $\f$ goes to infinity, then for $t\rightarrow \infty$, $\f 
\simeq  \sqrt{a}\ln (t)$, $\rho+3p\sim (1+3\nu)/t^{3a (1 + \nu)}$ with $\nu=
p'(0)$, and, by (\ref{eq:ac}), $\ddot{R} /R \simeq a(a -1)/ t^2$ if  $3a (1 + 
\nu)>2$, which is true for $1 + 3\nu>0$, i. e., exactly when the strong energy 
condition $\rho + 3p >0$ holds for small densities. The expansion in this case 
accelerates.

b) {\it Potentials vanishing as a power at infinity}. For some $b> 0$, let
$\C = [4/(b+4)]^2 (\sqrt{b}/ \f)^b \left[1+{\rm o}(1)\right]$ as $\f\rightarrow
+\infty$, and this asymptotic formula may be differentiated. If at large times 
$\f$ goes to infinity , then for $t\rightarrow \infty$, $\f \simeq  \sqrt{b} \,
t^{2/(b+4)}$, $\rho+3p\sim (1+3\nu)\exp{[-3(1+ \nu)t^{4/(b+4)}]}$ with $\nu=
p'(0)$, and, by (\ref{eq:ac}), $\ddot{R} /R \simeq [4/(b+4)]^2 t^{-2/(b+4)}$, 
so that the expansion accelerates. 

These two types of potentials were the ones studied in \cite{q1} -- \cite{q2}.

To make the picture complete, we consider now the (not that interesting) 
exponential regime of expansion, i. e., solutions with $\rho_T^{\infty} > 0$. 

The large time behavior of the scalar field is described in this case by the 
damped anharmonic oscillator equation (\ref{eq:fa}). According to the Poincar$
\acute e$--Bendixson theory of autonomous 
dynamical systems on the plane (e. g.~\cite{rei}, 2.8), every  bounded 
non-closed phase trajectory has a limit trajectory which is either a critical 
point or an isolated closed orbit (limit cycle, periodic solution). However, 
equation (\ref{eq:fa}) has no periodic solutions at all. Indeed, it can be 
rewritten as
$$
\dot{\rho_\f}=-2\beta\dot{\f}^2,
$$
and the assumption that $\f(t)$ is periodic with some period $T>0$ implies
$$
\int_0^T\dot{\rho_\f}\,dt=0=-2\beta\int_0^T\dot{\f}^2\,dt,
$$
that is, $\dot\f(t)\equiv0$, because $\beta>0$. The derivative of the scalar 
field is always bounded, $\dot{\f}^2(t)<\rho_T(t_0)$ for $ t > t_0$. Therefore
if the scalar field is bounded, then the solution
goes to a critical point, $\lim_{t \rightarrow \infty} \dot{\f}(t)=0$,
$\lim_{t \rightarrow \infty} {\f(t)}=\f_{\infty},\,|\f_{\infty}|< +\infty$, $
V(\f_\infty)=\rho_T^{\infty} > 0$, and, by (\ref{eq:fa}), $d \C / d \f 
|_{\f=\f_{\infty}}=0$, $d^2 \C / d \f^2 |_{\f=\f_{\infty}} >0$. Again the 
field, and with it the whole system, evolves towards a finite stable critical 
point which now corresponds to a non-zero minimum of the potential. This is 
the only possibility for a system with a growing potential, $\lim_{\f
\rightarrow\pm \infty}\C=+\infty$, because under this condition all 
trajectories are obviously bounded. The system in such a case is dissipative, 
and its limit behavior is conveniently described in terms of the connected 
global attractor consisting of all the critical points and (heteroclinic) 
trajectories connecting the unstable critical points with the stable ones 
(~\cite{lad},~\cite{ha}).

It remains thus to find out about the unbounded phase trajectories of 
(\ref{eq:fa}),  if they exist at all, that is, under the condition that $\C$ 
is bounded at least at one of the infinities. As explained, the unbounded 
trajectory means that $\lim_{t \rightarrow \infty}\f(t)=\pm\infty$; to be 
precise, let us speak about the positive infinity. If the limit of the 
potential, $\lim_{\f\rightarrow+\infty}\C=V_\infty<+\infty$ exists, then 
clearly $\lim_{t \rightarrow \infty} \dot{\f}(t)=0$, $V_\infty=\rho_T^{\infty}
$, so that the whole system once again tends to a stable fixed point, with 
just an infinite limit value of the scalar field. 

The exceptions of evolution towards a critical point thus could possibly  be 
provided only by the potentials which are bounded but have no limit at 
infinity (it looks rather peculiar from the physical point of view, unless 
they are periodic). Our survey of the results pertinent to this 
case so far has shown that one cannot completely rule 
out the solutions which undergo rather strange damped oscillations 
characterized by the following: $\lim_{t \rightarrow \infty}\f(t)=\pm\infty$, 
$\dot\f(t)$ is bounded and has no limit (in particular, $\lim_{t \rightarrow 
\infty} \dot{\f(t)}\not=0$), $\lim_{t \rightarrow \infty}\rho_\f(t)=
\rho_T^{\infty}>0$, and
$$
\int_\tau^\infty\dot{\f}^2\,dt<+\infty,\qquad \tau\geq t_0
$$
If these exist at all, the relevance of such regimes to the Universe appears to
be questionable at the very least.

Let us now give a list of the most significant facts regarding the large time 
behavior of the cosmological expansion we have so far established.

1. The matter density  decreases to zero, the total energy density decreases to
a (nonnegative) constant, the scalar field energy density is non-increasing and
goes to the same constant
as the total density, always.

2. Except possibly for such potentials $\C$ that are bounded at infinity but 
have no limit there, the dynamical system always evolves towards one of its 
critical (fixed) points with either finite or infinite limit value $\f_\infty$ 
of the scalar field; the time derivative of the field goes to zero in both 
cases; if the limit value of the field is finite, the potential has a minimum 
at it.

3. The expansion regime at large times is either exponential or subexponential
with $\dot{R}/R\rightarrow0$; in the (typical) case of evolution to a fixed 
point the former occurs if $V(\f_\infty)>0$, the latter if  $V (\f_\infty)=0$.

4. The subexponential expansion may be accelerating, that is, the scalar field
can dominate "normal" matter ($\rho + 3p > 0$), provided that the scalar field
tends to infinity and the potential has a zero limit $V(\f_\infty)=
V(\infty)=0$ and certain asymptotic behavior there.

Properties 1 -- 3 of the cosmological expansions are hardly surprising, 
however, all these results are now firmly and unambiguously established under 
general conditions of a clear physical origin.

Given the potential, the choice between the described possibilities is made by
the initial conditions, so for the same potential the final stage of the 
expansion might be different depending on what happened in the early Universe.
Let us illustrate this by some examples.

If $\C$ is a potential well, i. e.,  it has a single minimum at some $\f=
\f_*$ and goes (however slowly!) to infinity on both sides of it, all the 
solutions, disregarding the initial conditions they stem from, tend to the 
single (stable) equilibrium at $\f_*$; depending on whether $V(\f_*)=0$ 
or not, it is  either a subexponential regime or an exponential one.

A much more sophisticated picture arises with, say, the potential plotted in 
the Figure.
It has a zero minimum at $\f=\f_2$, a non-zero minimum at $\f=\f_4$, two 
unstable equilibria (maxima) at $\f=\f_{1,3},\,\,\f_1<\f_2<\f_3<\f_4$, goes to
infinity
as $\f\rightarrow +\infty$ but tends to zero when $\f\rightarrow -\infty$. 
Both exponential ($\f_\infty=\f_4$) and subexponential ($\f_\infty=\f_2,
\, -\infty$) regimes are possible, depending on how the expansion starts; the 
latter regime with $\f_\infty=-\infty$ can be accompanied by acceleration even 
if $\rho + 3p > 0$, provided that $\C$ has the right asymptotic behavior at $
\f\rightarrow -\infty$. The exponential growth occurs, for instance, when $
\f_0>\f_3$ and $\rho_\f(t_0)< V (\f_3)$; subexponential expansion without 
acceleration happens for the initial conditions with $\f_1<\f_0<\f_3$ and $
\rho_\f(t_0)< V (\f_1)< V (\f_3)$ ($\f\rightarrow\f_2$); subexponential
regime with possible acceleration takes place when $\f_0<\f_1$ and $\rho_\f(t_0
) < V (\f_1)$, when $\f\rightarrow-\infty$.

Our final remark is about the closed Universe, $k=1$. The only thing we need to
validate our results in this case is that the expansion never stops (and maybe 
turns into contraction), i. e., that $\dot{R} > 0$ for any $t>t_0$. By (0.6), 
we are speaking about solutions for which $\rho_T- 3/R^2$ is positive for all $
t>t_0$ as soon as it is positive at the initial moment $t=t_0$; for any such 
"infinite expansion solution", our results are true. There are many conditions 
under which these solutions exist. For instance, if $\C\geq V_{min}>0$, 
then, since $\rho_T>\C$ by definition, any solution starting with $R_0>R_{min}=
\sqrt{3/ V_{min}}$ is an infinite expansion solution. The other condition 
which does not depend on the potential at all is derived from the fact that, by
the same definition, $\rho_T>8\pi\rho$. The energy conservation equation (0.10)
shows that at large times $\rho\sim R^{-3(1+\nu)}$, $\nu\equiv p'(0)$. If 
$p'(0)<-1/3$, then $-3(1+\nu)<2$, and any solution with large enough 
"starting radius" $R_0$ satisfies $\rho_T>3/R^2$ and hence is an infinite 
expansion solution, etc..

Therefore it seems that acceleration of the Universe is a fairly general 
behavior. What is surprising in light of recent supernovae results \cite{sn1a1}
is that the period of acceleration is just starting around the present time. 
One would have to fine-tune the parameters of our potentials to account for 
this. This rather unfortunate situation would be resolved if we had an 
appropriate potential which came from more fundamental physics.

A paper with similar results for the technically much more complicated case of 
scalar-tensor theories of gravity with non-minimal coupling of the scalar 
field, is now in preparation.
\vskip5mm
\centerline{ \epsfxsize=0.7\hsize \epsffile{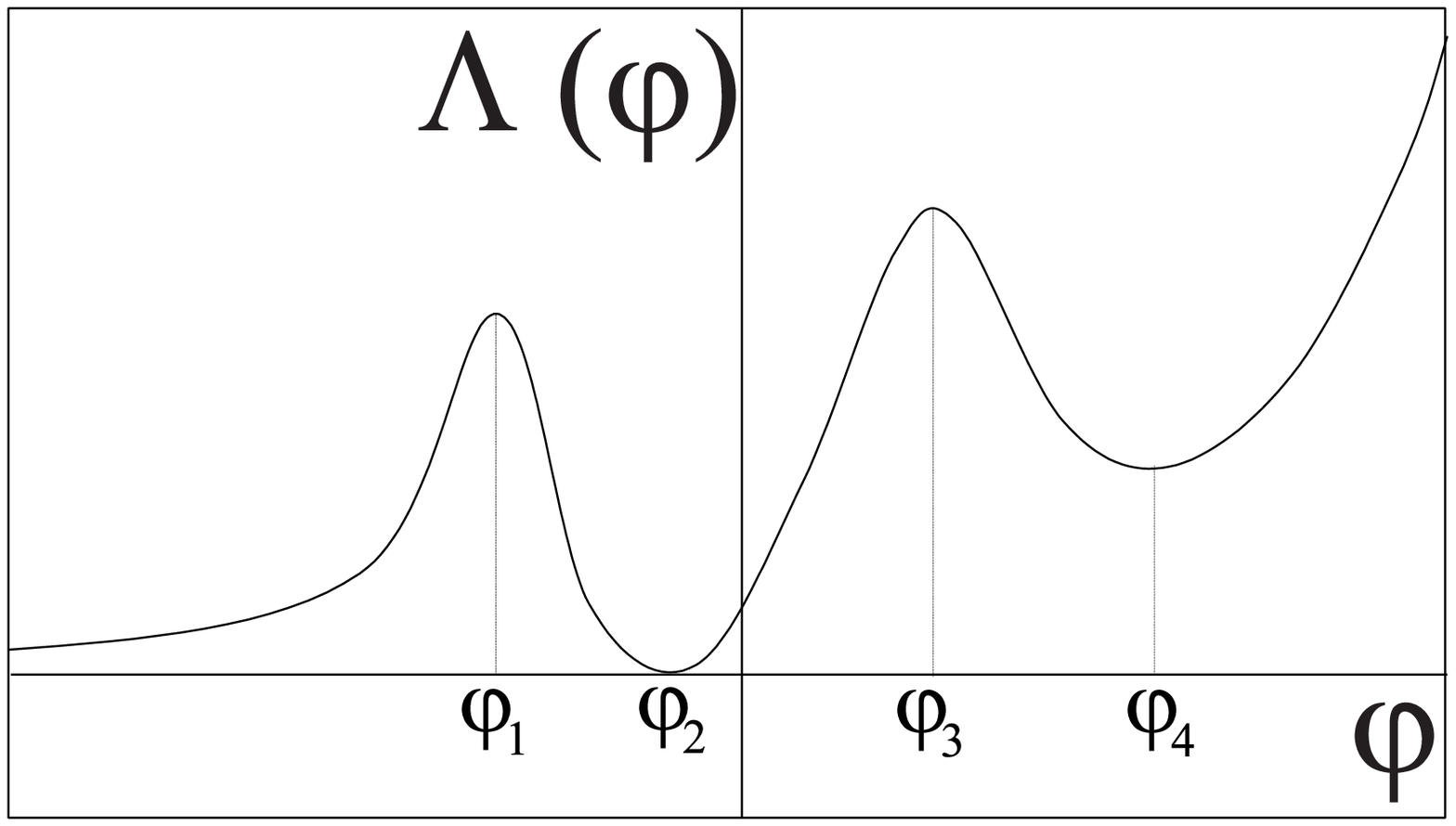}}

\section*{Acknowledgments}

This work was supported by NASA grant NAS 8-39225 to Gravity Probe~B. We are 
grateful to R.V.Wagoner for many valuable comments and to the Gravity Probe B 
Theory Group for fruitful discussions. We would also like to thank Manuel E. 
Cid and Ricardo D\'{\i}az D\'{\i}az for their help with the figure.

\end{document}